
\def\"#1{{\accent"7F #1\penalty10000\hskip 0pt plus 0pt}}

\newskip\ttglue
\newfam\sffam
%
%
\font\sixrm=cmr6
\font\sixi=cmmi6
\font\sixsy=cmsy6
\font\sixbf=cmbx6
%
\font\eightrm=cmr8
\font\eighti=cmmi8
\font\eightsy=cmsy8
\font\eighttt=cmtt8
\font\eightit=cmti8
\font\eightsl=cmsl8
\skewchar\eighti='177
\skewchar\eightsy='60
\hyphenchar\eighttt=-1     
\font\eightsf=cmss8
\font\eightssq=cmssq8
%
\font\ninerm=cmr9
\font\ninei=cmmi9
\font\ninesy=cmsy9

\font\ninebf=cmbx9

\font\ninesf=cmss8    scaled \magstephalf

\skewchar\ninei='177
\skewchar\ninesy='60
%
\font\tensf=cmss10
\font\tenssq=cmssq8 scaled \magstep1
\font\tensc=cmcsc10
%
\font\elevenrm=cmr10 scaled \magstephalf
\font\eleveni=cmmi10 scaled \magstephalf
\font\elevensy=cmsy10 scaled \magstephalf
\font\elevensc=cmcsc10 scaled \magstephalf
\font\elevenex=cmex10 scaled \magstephalf
\font\elevenbf=cmbx10 scaled \magstephalf
\font\elevensl=cmsl10 scaled \magstephalf
\font\eleventt=cmtt10 scaled \magstephalf
\font\elevenit=cmti10 scaled \magstephalf

\font\elevensc=cmcsc10 scaled \magstephalf

\skewchar\eleveni='177
\skewchar\elevensy='60
\hyphenchar\eleventt=-1
%
\font\twelverm=cmr12
\font\twelvei=cmmi12
\font\twelvesy=cmsy10 scaled \magstep1
\font\twelvesc=cmcsc10 scaled \magstep1
\font\twelveex=cmex10 scaled \magstep1
\font\twelvebf=cmbx12
\font\twelvesl=cmsl12
\font\twelvett=cmtt12
\font\twelveit=cmti12
\font\twelvesf=cmss12
\font\twelvessq=cmssq8 scaled \magstep2
\skewchar\twelvei='177
\skewchar\twelvesy='60
\hyphenchar\twelvett=-1
%
\font\fourteenrm=cmr12 scaled \magstep1
\font\fourteeni=cmmi10 scaled \magstep2
\font\fourteenit=cmti12 scaled \magstep1
\font\fourteensy=cmsy10 scaled \magstep2
\font\fourteensc=cmcsc10 scaled \magstep2
\font\fourteenbf=cmbx12 scaled \magstep1
\font\fourteensl=cmsl12 scaled \magstep2
\font\fourteentt=cmtt12 scaled \magstep1
\font\fourteensf=cmss12 scaled \magstep1
\skewchar\fourteeni='177
\skewchar\fourteensy='60
\hyphenchar\eighttt=-1
%
%
%
%
\def\sixpoint{\def\rm{\fam0\sixrm}%
   \textfont0=\sixrm  \scriptfont0=\fiverm \scriptscriptfont0=\fiverm
   \textfont1=\sixi \scriptfont1=\fivei \scriptscriptfont1=\fivei
   \textfont2=\sixsy  \scriptfont2=\fivesy
	\scriptscriptfont2=\fivesy
   \textfont\bffam=\sixbf  \scriptfont\bffam=\fivebf
      \scriptscriptfont\bffam=\fivebf  \def\bf{\fam\bffam\sixbf}%
   \setbox\strutbox=\hbox{\vrule height7.5pt depth2.5pt width0pt}%
   \rm}
%
\def\eightpoint{\def\rm{\fam0\eightrm}%
   \textfont0=\eightrm  \scriptfont0=\sixrm  \scriptscriptfont0=\fiverm
   \textfont1=\eighti \scriptfont1=\sixi \scriptscriptfont1=\fivei
   \textfont2=\eightsy  \scriptfont2=\sixsy  \scriptscriptfont2=\fivesy
   \textfont3=\eightex  \scriptfont3=\eightex
   \scriptscriptfont3=\eightex
   \textfont\itfam=\eightit  \def\it{\fam\itfam\eightit}%
   \textfont\slfam=\eightsl  \def\sl{\fam\slfam\eightsl}%
   \textfont\ttfam=\eighttt  \def\tt{\fam\ttfam\eighttt}%
   \textfont\bffam=\eightbf  \scriptfont\bffam=\sixbf
   \scriptscriptfont\bffam=\fivebf  \def\bf{\fam\bffam\eightbf}%
   \textfont\sffam=\eightsf  \def\sf{\fam\sffam\eightsf}%
   \def\sc{\eightsc}
   \def\ssq{\eightssq}
   \tt \ttglue=.4em plus.25em minus.15em
   \setbox\strutbox=\hbox{\vrule height7.5pt depth2.5pt width0pt}%
   \rm}
%
\def\tenpoint{\def\rm{\fam0\tenrm}%
   \textfont0=\tenrm  \scriptfont0=\sevenrm  \scriptscriptfont0=\fiverm
   \textfont1=\teni \scriptfont1=\seveni \scriptscriptfont1=\fivei
   \textfont2=\tensy  \scriptfont2=\sevensy  \scriptscriptfont2=\fivesy
   \textfont3=\tenex  \scriptfont3=\tenex  \scriptscriptfont3=\tenex
   \textfont\itfam=\tenit  \def\it{\fam\itfam\tenit}%
   \textfont\slfam=\tensl  \def\sl{\fam\slfam\tensl}%
   \textfont\ttfam=\tentt  \def\tt{\fam\ttfam\tentt}%
   \textfont\bffam=\tenbf  \scriptfont\bffam=\sevenbf
   \scriptscriptfont\bffam=\fivebf  \def\bf{\fam\bffam\tenbf}%
   \textfont\bffam=\tenbf  \scriptfont\bffam=\eightsf
   \scriptscriptfont\bffam=\eightsf  \def\sf{\fam\sffam\tensf}%
   \tt \ttglue=.5em plus.25em minus.15em
   \def\sc{\tensc}
   \def\ssq{\tenssq}
   \def\tr{\tentr}
   \setbox\strutbox=\hbox{\vrule height8.5pt depth3.5pt width0pt}%
	       \rm}
%
\def\elevenpoint{\def\rm{\fam0\elevenrm}%
    \textfont0=\elevenrm \scriptfont0=\sevenrm \scriptscriptfont0=\fiverm
    \textfont1=\eleveni  \scriptfont1=\seveni \scriptscriptfont1=\fivei
    \textfont2=\elevensy \scriptfont2=\sevensy \scriptscriptfont2=\fivesy
    \textfont3=\elevenex \scriptfont3=\elevenex \scriptscriptfont3=\elevenex
    \textfont\itfam=\elevenit  \def\it{\fam\itfam\elevenit}%
    \textfont\slfam=\elevensl  \def\sl{\fam\slfam\elevensl}%
    \textfont\ttfam=\eleventt  \def\tt{\fam\ttfam\eleventt}%
    \textfont\bffam=\elevenbf  \scriptfont\bffam=\sevenbf
    \scriptscriptfont\bffam=\fivebf  \def\bf{\fam\bffam\elevenbf}%
    \tt \ttglue=.5em plus.25em minus.15em
    \def\sc{\elevensc}
    \def\tr{\eleventr}
    \setbox\strutbox=\hbox{\vrule height11pt depth3.3pt width0pt}%
    \rm}
%
\def\twelvepoint{\def\rm{\fam0\twelverm}%
   \textfont0=\twelverm \scriptfont0=\ninerm \scriptscriptfont0=\sevenrm
   \textfont1=\twelvei     \scriptfont1=\ninei \scriptscriptfont1=\seveni
   \textfont2=\twelvesy \scriptfont2=\ninesy \scriptscriptfont2=\sevensy
   \textfont3=\twelveex \scriptfont3=\twelveex \scriptscriptfont3=\twelveex
   \textfont\itfam=\twelveit  \def\it{\fam\itfam\twelveit}%
   \textfont\slfam=\twelvesl  \def\sl{\fam\slfam\twelvesl}%
   \textfont\ttfam=\twelvett  \def\tt{\fam\ttfam\twelvett}%
   \textfont\bffam=\twelvebf  \scriptfont\bffam=\ninebf
   \scriptscriptfont\bffam=\sevenbf  \def\bf{\fam\bffam\twelvebf}%
   \textfont\sffam=\twelvesf  \scriptfont\sffam=\ninesf
   \scriptscriptfont\sffam=\eightsf  \def\sf{\fam\sffam\twelvesf}%
   \tt \ttglue=.5em plus.25em minus.15em
   \def\sc{\twelvesc}
   \def\ssq{\twelvessq}
   \def\tr{\twelvetr}
   \setbox\strutbox=\hbox{\vrule height10pt depth4pt width0pt}%
    \rm}
%
\def\fourteenpoint{\def\rm{\fam0\fourteenrm}
   \textfont0=\fourteenrm  \scriptfont0=\twelverm  \scriptscriptfont0=\tenrm
   \textfont1=\fourteeni \scriptfont1=\twelvei \scriptscriptfont1=\teni
   \textfont2=\fourteensy  \scriptfont2=\twelvesy  \scriptscriptfont2=\tensy
   \textfont3=\twelveex    \scriptfont3=\twelveex
   \scriptscriptfont3=\twelveex
   \textfont\itfam=\fourteenit  \def\it{\fam\itfam\fourteenit}%
   \textfont\slfam=\fourteensl     \def\sl{\fam\slfam\fourteensl}%
   \textfont\ttfam=\fourteentt  \def\tt{\fam\ttfam\fourteentt}%
   \textfont\bffam=\fourteenbf     \scriptfont\bffam=\twelvebf
   \scriptscriptfont\bffam=\tenbf  \def\bf{\fam\bffam\fourteenbf}%
   \tt \ttglue=.5em plus.25em minus.15em
   \def\sc{\fourteensc}
   \def\sf{\fourteensf}
   \def\tr{\fourteentr}
   \setbox\strutbox=\hbox{\vrule height11.5ptdepth4.5pt width0pt}%
    \rm}
%

%
\twelvepoint

\vsize=9.2in
\hsize=6.5in

\clubpenalty=9000
\widowpenalty=9000

\font\twelverm=cmr12
\footline={\hss\twelverm\folio\hss}
\def\makefootline{{\baselineskip=40pt \line{\the\footline}}}
\def\newpage{\vfill\eject}
\def~{\penalty 10000 \hskip0.15em}
\def\slash#1{\setbox0=\hbox{$#1$}\setbox1=\hbox{$/$}\if\wd0>\wd1
   \copy0\kern-\wd0\hbox to
   \wd0{\hss\kern.1em\box1\hss}\else\copy1\kern-\wd1\hbox to
   \wd1{\hss\box0\kern.1em\hss}\fi}

\def\leaderfill#1{\leaders \hbox to 1em {\hss#1\hss}\hfill}

\hfill RUB--TPII--56/93
\vskip0.5\baselineskip
\hfill December, 1993 \

\vskip3\baselineskip
%
%
%

%

\twelvepoint
\baselineskip=19pt

\def\makefootline{\baselineskip=40pt \line{\the\footline} }

%

\def\Re{{\it Re}\,}
\def\Im{{\it Im}\,}
\def\sign{{\rm sign}}

%
%
%
%
\def\NJLLAG{1}
\def\NEWLAG{2}
\def\CHIRCIR{3}
\def\NEWGENp{4}
\def\ACTION{5}
\def\ACTa{10a}
\def\ACTb{10b}
\def\SPLOG{6}
\def\DIRAC{7}
\def\DIRACEL{11}
\def\HAMIL{8}
\def\DEFU{9}
\def\RDEF{13}
\def\DIRFUL{15}
\def\MESF{12}
\def\DIRFULp{14}
\def\OMHER{16}
\def\DIRCON{17}
\def\OMQ{18}
\def\NEWGEN{19}
\def\REIM{20}
\def\REIMa{20a}
\def\REIMb{20b}
\def\CURR{21}
\def\PROP{22}
\def\REGACTIO{23}
\def\HEDANZ{24}
\def\OPEXP{25}
\def\WDEF{26}
\def\WI{27}
\def\WIa{27a}
\def\WIbp{27b'}
\def\WIcp{27c'}
\def\WIb{27b}
\def\WIc{27c}
\def\WId{27d}
\def\VDEF{28}
\def\VDEFa{28'}
\def\TIMESP{29}
\def\CURRNR{30}
\def\EXPANSp{31}
\def\JFULL{32}

\def\FFDEFa{33a}
\def\FFDEFb{33b}
\def\FFDEF{33}
\def\FFSV{34}
\def\ELSCAL{35}
\def\EVCUR{36}
\def\OMR{37}
\def\ELVEC{38}
\def\MSCURa{39}
\def\MSCUR{40}
\def\MAGSCa{41}
\def\MAGSCb{42}
\def\MAGSC{43}
\def\MVCURp{44a}
\def\MVCUR{44b}
\def\MVCURap{45}
\def\MAGVEC{46}
\def\FFPNa{47a}
\def\FFPNb{47b}

\def\ELRAD{48}
\def\MAGMOM{49}

\def\EUCLID{A.1}
\def\APPAa{A.2}
\def\APPAaa{A.2a}
\def\APPAab{A.2b}
\def\APPAb{A.3}
\def\APPAc{A.4}
\def\APPAd{A.5}
\def\APPAe{A.6'}
\def\APPAf{A.6}
\def\APPAg{A.7}
\def\APPAh{A.8}
\def\APPAi{A.9}

\def\SKYR{1}
\def\NJL{2}
\def\EGUCHI{3}
\def\DP{4}
\def\DPP{5}
\def\RW{6}
\def\HEDGE{7}
\def\MESSECa{8}
\def\MESSECb{9}
\def\BARSECa{10}
\def\BALL{11}
\def\SRAL{12}
\def\GGGRW{13}
\def\WAKAM{14}
\def\GGG{15}
\def\RIPKA{16}
\def\SCHW{17}
\def\CORRa{18}
\def\CORRb{19}


\vskip1cm

\vskip\baselineskip
{\fourteenpoint
\bf\centerline {Electromagnetic nucleon properties and quark sea
polarization}
\vskip10pt
\bf\centerline {in the Nambu-Jona-Lasinio model}
}
\vskip1.5cm

{\sc
\centerline{A.Z.~G\'orski$^{1,2}$, \ Chr.V.~Christov$^1$,
\ K.~Goeke$^1$}  }
\vskip1.0cm

{\tenpoint\it
\item{$^1$} Institute f\"ur Theoretische Physik II,
Ruhr-Universit\"at Bochum, D-44780 Bochum, Germany

\item{$^2$} Institute of Nuclear Physics, Radzikowskiego 152, 31-342
Cracow, Poland  }

\vskip4\baselineskip

\centerline{ABSTRACT}
\vskip0.5\baselineskip

 In this paper we present the derivation as well
as the numerical results for all electromagnetic form factors
of the nucleon within the
semibosonized Nambu--Jona-Lasinio (chiral quark soliton) model.
Other observables, namely the nucleon mean squared radii, the
magnetic moments and the nucleon--$\Delta$ splitting
are also computed.
The calculation has been done taking into account the
quark sea polarization effects.
The final results, including rotational $1/N_c$ corrections,
are compared with the existent experimental
data and they are found to be in a good agreement for the
constituent quark mass of $400$--$420\ MeV$.

\newpage

\centerline{\bf 1. Introduction }
\vskip\baselineskip

 In the last years the most challenging problem in elementary particle
physics seems to be solution of QCD in the low energy region.
The main difficulties are due to the non-perturbative effects caused by
growing effective coupling constant of the fundamental theory
in the low energy limit.
This prevents one from using the well-known main tool of theoretical
physics --- the perturbation theory.
Because of this, the most intriguing features of QCD -- confinement and
chiral symmetry breaking -- still remain conceptual and practical problems.

 The above mentioned obstacles have initiated an increasing interest
among the physicists in non-perturbative methods and effective
low-energy models of hadrons.
The effective models are expected to mimic the behaviour of QCD at
energies below $\sim 1~GeV$ (confinement and/or chiral symmetry
breaking) and to reproduce experimental data in this region.
In principle, those models could be related to QCD by
integrating out gluonic fields and reparameterizing fermionic
degrees of freedom (the bosonization procedure).
In effect, the physically observed particles should appear
as fundamental fields of the model.

 One of the first models of this type that has become
very popular was the Skyrme model [\SKYR].
It is an example of a fully bosonized effective model.
During many years of intensive investigations the Skyrme model
has proven to be theoretically quite interesting, though
comparison with experiment has not always been good.

 The simplest purely fermionic Lorentz invariant model with spontaneous
chiral symmetry breaking is the Nambu--Jona-Lasinio (NJL)
model [\NJL]
that has followed the example of the BCS model of superconductivity.
It contains chirally invariant local four-fermion interaction terms
and it is non-renormalizable in 4-dimensional space-time. The NJL
Lagrangian in its simplest $SU(2)$ form has the following structure:

$$ {\cal L}_{NJL} = \bar q \, (i \slash\partial - m_0)\, q +
{1\over 2} G \,
\,[\,(\bar q q )^2 + (\bar q i \vec\tau \gamma_{5} q )^2\,                ]
\ ,  \eqno(\NJLLAG) $$

\noindent where $G$ is the coupling constant, $m_0$ is the current
quark mass and $\vec\tau$ are the Pauli matrices in the isospin
space. Usually $up$ and $down$ quarks are assumed as degenerated
in mass.

 The NJL model is generally solved after applying the well known
bosonization procedure
following Eguchi [\EGUCHI] to arrive at the model expressed in terms of
physically observed particles' fields
$\sigma({\vec x}), {\vec\pi}({\vec x})$:

$$ {\cal L}_{NJL}' = \bar q \  [ \ i \gamma^{\mu}\partial_{\mu} -
g (\sigma + i\, {\vec\pi \vec\tau}\, \gamma_5 )\ ] \, q -
{1\over 2} \mu^2 (\sigma^2 + {\vec\pi}^2) +
(m_0\, \mu^2 / g) \, \sigma
\ . \eqno(\NEWLAG) $$

\noindent The new coupling constant $\mu^2$ is related to
the initial $G$ by: $G=g^2/\mu^2$. Here, $g$ is the physical
pion--quark coupling constant implying that $\vec\pi$
are the physical pion fields.
The meson fields are constrained to the chiral circle
to reproduce the pion decay constant:

$$ \sigma^2({\vec x}) + {\vec\pi}^2({\vec x}) = f^2_{\pi}
\ , \eqno(\CHIRCIR) $$

\noindent where $f_{\pi}=93~MeV$ is the pion decay constant.

In fact, an equivalent effective chiral quark meson theory [\DP,\DPP] can be
derived from the instanton model of the QCD vacuum.

In the chiral quark soliton model [\DPP-\HEDGE] based on the lagrangean
(\NEWLAG) (frequently reffered simply as NJL model) the baryons appear as a
bound state of $N_c$ valence quarks coupled to the polarized Dirac sea
($\overline qq$). Operationally the baryon sector of the model is solved in
two steps. In the first step, motivated by the large $N_c$ limit, a static
localized
solution (soliton) is found. It is done by solving the equations of motion
in an iterative selfconsistent procedure assuming that the $\sigma$ and
$\vec\pi$ fields have hedgehog structure. However, this hedgehog soliton
does not preserve the spin and isospin. In order to describe the nucleon
properties one  needs nucleon states with the proper spin and isospin
numbers. To this end the classical solution is semiclassically quantized
making use of the rotational zero modes. In fact, such a cranking procedure
was elaborated in [\DPP]. Recently, within this scheme including Dirac sea
polarization effects quite successful calculations for the nucleon moment
of inertia [\GGGRW], magnetic moments [\WAKAM] and nucleon electric form
factors [\GGG] have been reported.

 There are several reasons that the model described by (\ACTION)
is considered as one of the most promising effective theories describing
low energy QCD phenomena.
First, the model is the simplest quark model which provides mechanism
for spontaneous breaking of the chiral symmetry --- the very basic
feature of QCD.
The mesons appear as excited states of quark-antiquark pairs.
The philosophy behind this approach is based on the hypothesis
that the chiral symmetry breaking and confinement are
relatively independent mechanisms and they can be investigated
separately [\DPP].

Second, the self-consistent solitonic solution ({\it hedgehog})
has been found to exist in the large $N_c$ limit of the model within
the physically acceptable range of parameters [\HEDGE].
The chiral soliton
provides a good description of the nucleon and gives us possibility
to take into account vacuum polarization effects from
the Dirac quark sea.
Calculations done in the semiclassical quantization procedure
yield quite reasonable results for the vacuum
and nucleon sectors [\MESSECa-\BARSECa].
In particular, quantities like the nucleon radii, $\Sigma$--terms,
axial vector coupling constant, $\Delta$--nucleon splitting,
splittings within the octet and decuplet are well reproduced.

 Third, there are various hints that the NJL-type Lagrangian can
be obtained [\DP,\DPP,\BALL,\SRAL] as a long wavelength limit of QCD.
It should be stressed that the large $N_c$ limit
plays a prominent role in those considerations.

 The aim of this paper is to compute the electromagnetic nucleon form
factors within the $SU(2)$ chiral quark soliton model based on the
semibosonized NJL-type lagrangean
with the vacuum polarization effects taken into account.
The calculations will include the rotational $1/N_c$ corrections,
which really have been shown to be important for the
isovector magnetic moment. Calculations of this sort
give us possibility to calculate all electromagnetic
properties of nucleons, as well as such quantities like the
electric mean squared radii, the magnetic moments, the
nucleon--$\Delta$ energy splitting, and the electric and magnetic
charge distributions.

 The paper is organized as follows: we begin in Sec.~2 with
formulation of the regularized model, introduction of the
rotational zero modes and we derive expression
for the electromagnetic current treating the angular velocity
as a perturbation.
In Section 3 we derive the analytical expressions for the form
factors including terms up to the linear order in angular
velocity.
Section 4 is devoted to presentation of numerical results. Summary and
discussion is given in the last section.

\vskip\baselineskip
\vskip\baselineskip
\centerline{\bf 2. The regularized action and current}
\vskip\baselineskip

In Minkowski space the generating functional of the
model written in terms of the quark and meson fields,
$\sigma({\vec x}), {\vec\pi}({\vec x})$ has the form:

$$ Z_{NJL}' = \int {\cal D}\bar q {\cal D}q
{\cal D}\sigma {\cal D}{\vec \pi}
\exp\left[\, i \int d^4x {\cal L}_{NJL}' (x) \right] \ ,
\eqno(\NEWGENp) $$
The sources are not explicitly included.

Since the NJL model is a non-renormalizable theory,
a regularization scheme of an appropriate cut-off $\Lambda$ is needed to
make the theory finite. To this end  we will work in Euclidean space-time
(the conventions used are given in the Appendix).

After Wick rotation (transformation to the Euclidean space-time) integrating
out the quarks in (\NEWGENp)
one gets the effective action split in fermionic and mesonic parts:
$$
S_{eff} = S^F_{eff} + S^M_{eff} + S_{eff}^{br} \ ,\eqno(\ACTION)
$$
The fermionic part of the action includes the quark loop contribution in
the presence of the external meson fields ($\sigma, \vec\pi$):
$$
S^F_{eff} = -  {\rm Sp}\, \ln (\ {\slash D}_E\ ).
\eqno(\SPLOG)
$$
The Dirac operator $\slash D_E$
(in Euclidean space-time) has the form:
$$
{\slash D}_E =  \gamma_4 \ \left[\partial_4 - h(U)\right] .
\eqno(\DIRAC)
$$
with the one particle Hamiltonian operator $h_E$ defined as
$$
h_E(U) \equiv i \, \left[ -i\, \gamma_4 \gamma_k \partial_k +
\gamma_4 \, M U \right] \ .
\eqno(\HAMIL)
$$
The constituent quark mass $M=gf_{\pi}$ and
$$
U = {1\over f_{\pi}} \, (\sigma + i \gamma_5 \vec\tau \vec\pi)
\   \eqno(\DEFU)
$$
represents the mesonic fields.
It can be checked by inspection that the Hamiltonian is
Hermitian: $ (h_E)^{\dagger} = h_E $.

The mesonic part
$$
S^M_{eff} = {\mu^2 \over 2} \int d^4 x_{E}\,( \sigma^2 +
\vec\pi^2 )
\ ,\eqno(\ACTa)
$$
and
$$ S_{eff}^{br} = - {m_o \mu^2 \over g} \int d^4x_E \ \sigma \
,\eqno(\ACTb)
$$
includes an explicit symmetry breaking term (\ACTb).

In order to compute the electromagnetic current matrix elements we
include the electromagnetic coupling in a minimal way and
we take into account the rotational zero modes.

We start with the Dirac operator ${\slash D}_E$  with electromagnetic
potential $A_\mu$:
$$
{\slash D}_E =  \gamma_4 \ \left[
\partial_4 - h(U) - i A_4 \hat Q +
i \gamma_4 \gamma_k A_k \hat Q \right] \ ,
\eqno(\DIRACEL)
$$
The quark charge matrix
$\hat Q \equiv {1\over 6} \hat1 + {1\over2} \tau^3$.


In the next step, we introduce the rotating soliton:
$$
\tilde U(\vec x,x_4) \equiv R(x_4) \, \bar U(\vec x) \,
R^{\dagger}(x_4), \qquad \ R\in SU_2,\qquad R^{\dagger} R = \hat 1.
\eqno(\MESF)
$$
where the meson field $\bar U(\vec x)$ corresponds to the stationary point
(minimum) of the effective action (\ACTION).  In the Euclidean space-time
the rotation matrix $R(x_4)$, can be represented as:

$$ R(x_4) = \exp(-i x_4 \, \Omega) \ , \quad\quad
\Omega \equiv {1\over 2} \tau^A \, \Omega^A
\quad \hbox{and}\quad
\Omega_E = - i \ R^{\dagger} \, \dot{R}
\ ,  \eqno(\RDEF) $$

\noindent where as usually the summation convention over
repeated indices is assumed.

In body-fixed frame the Dirac operator of the rotating soliton
$\tilde {\slash D}_E(\bar U)$  can be expressed in the form:

$$ \tilde{\slash D}_E(\bar U) =  R^{\dagger} \
{\slash D}_E(\tilde U) \ R \ ,
\eqno(\DIRFULp) $$

\noindent where $\tilde{\slash D}_E$ is defined by:

$$    \tilde{\slash D}_E(\bar U) = \gamma_4 \left[ \partial_4 - h(\bar U) +
i \Omega - i A_4 \, R^{\dagger} \hat Q R +
i \gamma_4 \gamma_k A_k \, R^{\dagger} \hat Q R \right]
\ . \eqno(\DIRFUL) $$

\noindent The operator $\Omega$ is Hermitian in both,
Euclidean and Minkowski space-time:

$$ \Omega^{\dagger}_E = \Omega_E, \quad\quad
\Omega^{\dagger}_M = \Omega_M
\ . \eqno(\OMHER) $$

\noindent Taking into account (\EUCLID) and (\OMHER)
we can write down the Hermitian conjugate to (\DIRFUL):

$$ \,\tilde{\slash D}_E^{\dagger}(\bar U) =
\left[ - \partial_4 - h(\bar U)  - i \Omega + i A_4^{\dagger} \,
R^{\dagger} \hat Q R + i \gamma_4 \gamma_k A_k^{\dagger}
\,  R^{\dagger} \hat Q R \, \right]
\gamma_4
\ , \eqno(\DIRCON) $$

\noindent where the collective coordinate $\Omega_M$
is will be quantized  according to the canonical quantization rule,
which in Minkowski space-time are:

$$ \Omega^A \rightarrow {1\over I} \  \hat{T}^A \equiv
{1\over 2I} \  \tau^A \ .
\eqno(\OMQ) $$

\noindent Here $\hat{T}^A$ are the $SU(2)$ group generators.
The moment of inertia $I$ has been computed numerically in [\GGGRW].

Now, using the ansatz (\DIRFULp)  the generating functional (\NEWGENp)  can
be re-expressed in terms of the rotational matrix $R(t)$ in Euclidean
space-time as
$$
Z'_{NJL}[A_{\mu}] = \int
{\cal D}R(t) \ \exp \big\lbrack - S_{eff}[A_{\mu}, R(t), \Omega] \,
\big\rbrack \ ,
\eqno(\NEWGEN)
$$
\noindent where $\Omega$ is a function of $R\equiv R(t)$
(see eq. (\RDEF)).
 Having the explicit form of the Dirac operator (\DIRFUL) we can
go back to the action (\ACTION).
Since we treat the meson fields classically ({\it i.e.} at
the 0--loop level) the only non-trivial part of the action
is the fermionic part (\SPLOG).
By construction (see [\DPP,\HEDGE]) it is splitted
in a natural way into valence and Dirac sea parts.
The valence part is finite and it will not be regularized.
In this case the calculation is straightforward and
the detailed derivation can be found {\it e.g} in [\DPP].
Hence, from now on we will concentrate on the sea contribution
and we will restrict our discussion to that
part of the full action.

 As we can see from (\DIRFUL), (\DIRCON), the operator (\DIRFUL)
is non-Hermitian. Hence, in general,
the action (\SPLOG) will have the real and imaginary part.
To make clear distinction we write down both parts explicitly:

$$\openup 2\jot
\eqalignno{
S_{eff}^F &= \Re S_{eff}^F + \Im S_{eff}^F \ , &(\REIM) \cr
\Re S_{eff}^F &\equiv - {1\over2}\ {\rm Sp} \ln
[\,\, \tilde{\slash D}^{\dagger}
\, \, \tilde{\slash D}\,] \ , &(\REIMa) \cr
\Im S_{eff}^F &\equiv - {1\over2}\ {\rm Sp} \ln
[\,\,\tilde{\slash D}\, /
\, \,\tilde{\slash D}^{\dagger}\,]
\ , &(\REIMb) \cr
}  $$

\noindent The imaginary (anomalous) part is finite and does not need
regularization. In addition, any regularization of this part of the
action would break several important features of the model
related to the anomalous structure.
On the other hand, the real part is infinite
and a regularization is necessary.

 In order to compute the nucleon form factors we have to evaluate
the expectation value of the electromagnetic current $j^{em}_{\mu}$.
To this end we take into account the standard definition
given by:

$$ \langle j_{\mu}^{em}(0) \rangle \equiv
{1\over Z_0} \
{\delta Z'_{NJL}[A_{\mu}]\over \delta A_{\mu}(0)} \Bigg\vert_{A_{\mu}=0}
\equiv {1\over Z_0} \
\int {\cal D}R \ {\delta S_{eff} \over \delta
A_{\mu}(0)}\Bigg\vert_{A_{\mu}=0} \exp \lbrack -\, S_{eff} \rbrack
\ , \eqno(\CURR) $$

\vskip0.1\baselineskip
\noindent where the generating functional $Z'_{NJL}$ is given by
(\NEWGEN).

 As it was mentioned above the real part of the action diverges.
To calculate the expression (\CURR), we introduce
the Schwinger proper-time regularization of the action.
In general, for an operator ${\cal A}$
its proper-time regularized form reads [\SCHW]:

$$ \ln {\cal A} = - \lim_{\Lambda\to\infty} \,
\int^{\infty}_{1/\Lambda^2} {d\tau\over\tau}
e^{-\tau{\cal A}}
\ . \eqno(\PROP) $$

\noindent The Schwinger method is especially convenient for the
logarithmic expressions resulting from the fermionic determinant.
Applying (\PROP) to the action (\SPLOG) we obtain:

$$    S^F_{eff} = {N_c \over 2}  \sum_{n} \int d^4 x_{E}\,
Tr_{\gamma} Tr_{\tau} \int_{1 \over\Lambda^2}^\infty {d\tau \over\tau}
\,\psi_{n}^{\dagger} (x_{E})\,\, \exp[\,-\tau
\,\tilde{\slash D}\,^{\dagger} \,
\tilde{\slash D} \,] \ \psi_{n} (x_{E})
\ , \eqno(\REGACTIO)  $$

\noindent where $\psi_n(x)$ is any complete set of eigenfunctions.
 Actually, in our calculations we shell use for $\psi_n(x)$
eigenfunctions of the Hamiltonian $h$.
For the meson fields $\sigma, \vec\pi$ we substitute the
static selfconsistent hedgehog solution $\bar U(\vec x)$ of the
chiral soliton [\HEDGE] form:

$$ \sigma(x) = \sigma(r), \quad \quad \vec\pi(x) =
\hat r \, \pi(r)
\ , \eqno(\HEDANZ) $$

\noindent restricted to the chiral circle (\CHIRCIR).
Numerically, we will follow the method of Ripka and Kahana
[\RIPKA] and will use the eigenfunctions of the
Hamiltonian $h$ to compute all matrix elements
(see also Section 4).

 Now, our task is to evaluate the current (\CURR) with the
regularized action
(\REGACTIO). The difficulty lies in the fact that the expression
(\REGACTIO) contains rather complicated operator exponent in the
integrand. However, because afterwards we take derivative
over the fields $A_{\mu}$ and put $A_{\mu}=0$, the only nonzero
contribution will come from the part of the integrand linear
in $A_{\mu}$.
We shall treat the cranking parameter $\Omega \sim {1\over N_c}$
as a small perturbation that is consistent with the large $N_c$ limit
philosophy behind the NJL model.
Hence, we can neglect terms $\Omega^2$ and
higher\footnote{$^*$}{Actually, the terms $\sim \Omega^2$ must be taken
into account to evaluate the moment of inertia (see [\DPP,
\GGGRW] for details).}.
Also, we should be careful in performing the ${\cal D}R$
integration, as in general the operators $\hat Q$ and $R$
does not commute and one must use the proper time
ordering (see Sect.~4 and [\CORRa, \CORRb]).
 In the next step we expand the integrand in
(\REGACTIO) in terms of $A_{\mu}$ and $\Omega$.
To this end we apply the Feynman-Schwinger-Dyson (FSD)
expansion for the operator exponent:

$$ e^{{\cal A}+{\cal B}} = e^{\cal A} +
\int^1_0 d\alpha\,e^{\alpha {\cal A}}\ {\cal B}\ e^{(1-\alpha)
{\cal A}} +
\int^1_0 d\beta\,\int^{1-\beta}_0 d\alpha\,
e^{\alpha {\cal A}}\ {\cal B}\
e^{\beta {\cal A}}\ {\cal B}\
e^{(1-\alpha-\beta) {\cal A}} + ...
\ .\eqno(\OPEXP) $$

\noindent In our case, for the operator  ${\cal A}$ we substitute
the part of our operator exponent ($-\tau W$), where:

$$ W \equiv \  \tilde{\slash D}_E\,^{\dagger} \
\tilde{\slash D}_E
\ , \eqno(\WDEF) $$

\noindent that does not contain $A_{\mu}$ and $\Omega$.
This part of $W$ we will call $W_0$.
On the other hand, for the operator ${\cal B}$ we take the
remaining terms of the operator $W$ and we will call them
$W_1, W_2,...,W_5$. The first type of terms ($W_1$) is defined
as the sum of all terms in the exponent that are linear in
$A_{\mu}$, \ $W_2 \sim A_{\mu} \times \Omega$, \ $W_3 \sim
\Omega$, \ $W_4 \sim \Omega^2$ and $W_5 \sim A_{\mu}^2$.
 The $W_5$ contribution to (\CURR) is exactly zero, while
the $W_4$ contribution we neglect as being small (quadratic
in $\Omega$). Hence, the operator ${\cal B}$ will
include the terms $W_1$--$W_3$ only.
Finally, this gives us the following expressions for the
terms $W_0$--$W_3$ of the operator $W$ (\WDEF):

$$ \openup 2\jot \eqalignno{
W_0 =  &+\omega^2 + h^2 \ , &(\WIa) \cr
W_1 = &- (A_4 + A_4^{\dagger}) \  R^{\dagger} \hat Q R \, \omega +
(A_k - A_k^{\dagger}) \  \gamma_4 \gamma_k \, R^{\dagger} \hat Q R
\, \omega \  + \cr
&+ i\,  ( h \, A_4\,  R^{\dagger} \hat Q R -
A_4^{\dagger}\,  R^{\dagger} \hat Q R \, h) \  - \cr
&- ( h \,  \gamma_4 \gamma_k \,  A_k\,  R^{\dagger} \hat Q R +
\gamma_4 \gamma_k\,  A_k\,  R^{\dagger} \hat Q R \,  h )
\ , &(\WIbp) \cr
W_2 =  &- ( A_4\, \Omega\, R^{\dagger} \hat Q R +
A_4^{\dagger} \,  R^{\dagger} \hat Q R \, \Omega ) +
\gamma_4 \gamma_k\, ( A_k\, \Omega\, R^{\dagger} \hat Q R -
A_k^{\dagger} \,  R^{\dagger} \hat Q R \, \Omega ) \  + \cr
&+ A_4\,  [\Omega, R^{\dagger} \hat Q R] - \gamma_4 \gamma_k \,
A_k\, [\Omega, R^{\dagger} \hat Q R] \  , &(\WIcp) \cr
W_3 =  &+2\, \Omega\,  \omega - i\,  [ h, \Omega ] \  . &(\WId)
}  $$

\noindent Taking into account that the electromagnetic potential is
Hermitian we get for the $W_1$ and $W_2$ the simpler form:

$$ \openup 2\jot \eqalignno{
W_1 = &+ i\,  [ h, A_4\,  R^{\dagger} \hat Q R] - i \,
\{ h, \gamma_4 \gamma_k\,  A_k\,  R^{\dagger} \hat Q R \} \  - \cr
&-2 A_4\, R^{\dagger} \hat Q R\,  \omega \ , &(\WIb)  \cr
W_2 =  &- 2\,A_4\, \Omega \, R^{\dagger}\, \hat Q \, R
 \ . &(\WIc) \cr
}  $$

\noindent where $[\ ,\ ]$ and $\{\ ,\ \}$ denote commutators
and anticommutators, respectively.
Also, the sum of all perturbative terms that will
contribute the electromagnetic current (\CURR) we will
denote by $V$:

$$ V \equiv W_1 + W_2 + W_3 \ . \eqno(\VDEF) $$

\noindent In addition, the part of $V$ containing
(linear) $\omega$-terms will be called $\omega \, V_1 $
and the $\omega$-indepen\-dent part will be denoted by
$V_0$. Hence, the full perturbative term reads:

$$ V = V_0 \ \omega^0 + V_1 \ \omega^1  \ .
\eqno(\VDEFa) $$

 To obtain (\WI) we have computed the trace over the Euclidean
time in the $\omega$-space, the Fourier conjugate to $x_4$.
To this end we have substituted in (\REGACTIO):

$$ \int dx_4 \ f(\partial_{x_4}) \rightarrow
\int {d\omega\over 2\pi} \ f(+i\omega)
\ . \eqno(\TIMESP) $$

\noindent Also, the formulae (\APPAc--\APPAd) from
the Appendix have been used.

 Applying (\WI) and (\OPEXP) we should have in mind that
there are, in general, two types of terms in (\OPEXP):
linear and bilinear in ${\cal B}$.
Hence, the products of the terms (\WIb-\WId) can contribute as well.
Also, those products should be linear in $\Omega$ and $A_{\mu}$.
This implies that the only $W$-term contributing to the second order
FSD expansion is of the form: $W_1 \times W_3$ --- the first one
linear in the potential $A_{\mu}$ and the second one linear in the
cranking parameter $\Omega$.
As we will see later on, those terms will contribute to the
electric isovector and  magnetic isoscalar parts of the form
factors.

 As the above considerations were limited to the regularized
part of the action a comment concerning the non-regularized
parts is in order. In general, there are two types of non-regularized
contributions: the valence parts (for all form factors) and
the sea parts of both isoscalar form factors.
To compute these quantities we can start directly from the
non-regularized action (\SPLOG) as it was done in [\DPP].
The electromagnetic current reads:

$$ j_{\mu}(x) \equiv {S_{eff}[A_{\mu}]\over\delta A^{\mu}}
\Bigg\vert_{A_{\mu}=0}
= N_c \, {\rm Tr}_{\gamma} {\rm Tr}_{\tau}
 \sum_n \int {d\omega\over 2 \pi} \, \psi_n^{\dagger}(x) \,
{1\over \omega + ih - \Omega} \, R^{\dagger} Q R \,\, \gamma_4
\gamma_{\mu} \, \psi_n(x)
\ , \eqno(\CURRNR) $$

\noindent where the integral over $\omega$ stands for
the trace in Euclidean time according to (\TIMESP).
Further calculations can be easier done, without using
the expansion (\OPEXP), just expanding (\CURRNR)
in terms of (small) $\Omega$.
The same results can be obtained by imposing the
$\Lambda^2\to\infty$ limit on our regularized
expressions (using formulae (\APPAa) of Appendix).

 The second remark concerns our practical calculation of the
imaginary part of the action. One can notice that taking
reversed Hermicity (anti-Hermitian) electromagnetic
potential $A_{\mu}$ in (\WIbp, \WIcp) the imaginary terms
are moved to the real part.
With this trick and the $\Lambda^2\to\infty$ limit we can
calculate all form factors from (\WI) within our scheme
and not referring to the nonregularized action (\SPLOG).

 Now, taking into account (\CURR), (\OPEXP) and (\WI-\VDEF)
we can write down the operator exponent contributing to the
electromagnetic current as:

$$ \openup 3\jot   \eqalign{
&\int^{+\infty}_{-\infty} {d\omega\over 2\pi} \ \exp [-\tau W]
\ \simeq \
- \tau \int_{-\infty}^{+\infty} {d\omega\over 2\pi} \sum_n
e^{-\tau \omega^2} \langle n \vert V \vert n \rangle \, + \cr
&+ \tau^2 \int_{-\infty}^{+\infty} {d\omega\over 2\pi}
\sum_{m,n} e^{-\tau \omega^2} \
\vert \langle n\vert V \vert m \rangle \vert^2
\int^1_0 d\beta \int^{1-\beta}_0 d\alpha \
e^{-\tau(1-\beta)\epsilon^2_n} \
e^{-\tau \beta \epsilon^2_m} \ , \cr  }
 \eqno(\EXPANSp) $$

 \noindent where the terms quadratic and higher in $\Omega$
are neglected and because of the derivative in (\CURR)
only the terms linear in potentials are present.
To compute the functional (space-time) trace
we use the Hamiltonian (\HAMIL), like for the
 action (\REGACTIO). Here, the vectors
$\vert n \rangle$ denotes the eigenfunctions $\psi_n(x)$
corresponding to the eigenvalues $\epsilon_n$.

 We cannot perform the $\omega$-integrals directly as the
operator $V$ is $\omega$-dependent.
However, we can simplify (\EXPANSp) taking into account
that the term $\vert V_{mn}\vert^2$ is symmetric
in indices $m,n$ and performing trivial $\alpha$- and
$\beta$-integrals.
After that the integration over $\omega$ is performed and
we get for the electromagnetic current the following
expression:

$$ \openup 3\jot  \eqalign{
j_{em}^{\mu} &= { N_c\over 4\sqrt{\pi}}
\int_{1/\Lambda^2}^{\infty} { d\tau\over\sqrt{\tau} }
\ \Bigg\lbrace
\sum_n e^{-\tau \epsilon_n^2} \
\langle n \vert {\delta V_0 \over \delta A_{\mu}}
 \, \vert n \rangle \ + \cr
&+ \sum_{m,n}
{e^{-\tau \epsilon_n^2} -e^{-\tau \epsilon_m^2} \over
\epsilon_n^2 - \epsilon_m^2 }
\ \Big[ \
\langle n \vert V_0[A_{\mu}=0] \vert m \rangle
\langle m \vert {\delta V_0 \over \delta A_{\mu} } \vert n \rangle
\ + \cr
&+ {1\over 2\tau} \, \langle n \vert V_1 [A_{\mu}=0]
\vert m \rangle
\langle m \vert {\delta V_1  \over \delta A_{\mu} }
\vert n \rangle \ \Big]
\ \Bigg\rbrace
\ , \cr
}
\eqno(\JFULL) $$

\noindent where the notation $V_1 [A_{\mu}=0]$ means
the sum of all terms linear in $\omega$
which do not contain $A_{\mu}$.
The formula (\JFULL) together with (\WI)
will be used in the next Section to
obtain the electromagnetic form factors' formulae.

\vskip\baselineskip
\vskip\baselineskip
\centerline{\bf 3. Computing the form factors}
\vskip\baselineskip

In this point, applying the FSD expansion (\OPEXP) we are ready to compute
the electromagnetic form factors
directly from the regularized action (\REGACTIO) which includes
the fermion loop corrections.
The nucleon electromagnetic Sachs form factors are defined
by:

$$ \openup 2\jot
\eqalignno{
\langle N(p)\vert j_0^{em}(0)\vert N(p')\rangle &= G_E(q^2)
\ , &(\FFDEFa) \cr
\langle N(p)\vert j_i^{em}(0)\vert N(p')\rangle &=
{1\over 2{\cal M}_N} \ G_M(q^2) \ i \
\epsilon_{ikl} \, q^k \langle N \vert \sigma^l \vert N \rangle
\ , &(\FFDEFb) } $$

\noindent where $\vert N(p') \rangle$ is the nucleon state with
proper spin and isospin quantum numbers.
The electromagnetic current $j^{em}_{\mu}(x)$ is in the Minkowski
space-time and ${\cal M}_N$ is the nucleon mass.
It is clear from (\FFDEF) that to compute the electromagnetic
form factors we need the current evaluated in the previous Section.
 The isoscalar and isovector parts of the form factors are defined by:

$$ G_{E,M} \equiv {1\over 2}\ G_{E,M}^{T=0} + \hat T_3 \
G^{T=1}_{E,M}
\ . \eqno(\FFSV) $$

\noindent The Sachs form factors (\FFDEF) are appropriate for the
non-relativistic limit (${\vec q\,}^2 \ll {\cal M}_N$)
that is consistent with the large $N_c$ approximation.

The isoscalar form factors (electric and magnetic) come from the scalar part
of the quark current matrix $\hat Q$ ($\sim \hat 1$)
while the isovector form factors is from the triplet part
($\sim \tau^3$).
The terms linear in $\Omega$ contribute to the electric isovector
and to both magnetic form factors.
As can be easily checked directly from
(\WI) and (\JFULL), the contribution from the real part of the action
to both the isoscalar electric form factor and
the isoscalar magnetic one is exactly zero.
This means that these form factors originate from the
imaginary part of the action and they will not be regularized.
Table 1 lists terms contributing to a given form factor.
In the last column we explicitly indicate which
form factor should be regularized.

\vskip\baselineskip
\item{Table 1.} Terms of the effective action contributing to different
electromagnetic form factors.

$$\vbox{\offinterlineskip
\hrule height1pt
\halign{&\strut\hfil#\hfil\quad
&\hfil#\hfil\quad
&\quad\hfil#\hfil\quad
&\quad\hfil#\hfil\quad
&\quad\hfil#\hfil\
&\quad\hfil#\hfil\quad\ \
&\  \hfil#\hfil\quad\ \
&\ \  \hfil#\hfil\cr
\noalign{\vskip3pt}
& quantity & $A_{\mu}$ & $\Omega$ & $\hat Q$ &
$2^0$-order FSD & $V_{\omega}$ & regularized &\cr
\noalign{\vskip6pt}
\noalign{\hrule height0.5pt}
\noalign{\vskip6pt}
& $G_E^{T=0}$ & $A_4$ & $\Omega^0$ & $\sim \hat 1$ & {\sc no} &
{\sc no} & {\sc no} &\cr
\noalign{\vskip6pt}
& $G_E^{T=1}$ & $A_4$ & $\Omega^1$ & $\sim \tau^3$ & {\sc yes} &
{\sc yes} & {\sc yes} &\cr
\noalign{\vskip6pt}
& $G_M^{T=0}$ & $A_k$ & $\Omega^1$ & $\sim \hat 1$ & {\sc yes} &
{\sc no} & {\sc no} &\cr
\noalign{\vskip6pt}
& $G_M^{T=1}$ & $A_k$ & $\Omega^0 + \Omega^1$ &
$\sim \tau^3$ & {\sc no} + {\sc yes} &
{\sc no} & {\sc yes} + {\sc no} &\cr
\noalign{\vskip6pt}
}
\hrule height1pt  } $$
\vskip\baselineskip

The electric scalar form factor can be immediately extracted from
(\FFDEFa) and (\CURRNR). Since only the isoscalar part
of the operator $\hat Q$ contributes the rotational matrices $R(t)$
vanish from the action. Hence, the $D{\cal R}$ integration in
(\NEWGEN) becomes trivial.
Also, in this case the contribution from the terms linear in
$\Omega$ is exactly equal to zero. Finally we arrive at:

$$\openup3\jot
\eqalign{
G_E^{T=0} (q^2) = &\int d^3 x\ e^{-i \vec q \vec x}
\ {N_c\over 3}
\ \Bigg\lbrace \sum_{n\in VAL} \psi_n^{\dagger}(\vec x) \,
\psi_n(\vec x)\ - \cr
&- \ \sum_{n\in ALL} \ {1\over 2}\ {\rm sign}(\epsilon_n)
\ \psi_n^{\dagger}(\vec x) \, \psi_n(\vec x)
\ \Bigg\rbrace \ , \cr }
\eqno(\ELSCAL) $$

\noindent where {\sc val} denotes the valence whereas {\sc all}
stands for all eigenstates of the Hamiltonian including the valence
one as well. The first term in eq.(\ELSCAL) comes from the valence quarks
and the second term is due to the quark vacuum polarization.
As has been stated previously, the same result can be obtained
by substituting anti-Hermitian fields $A_{\mu}$ in (\WIbp),
using (\JFULL), and taking the limit $\Lambda^2\to\infty$.

 As the next step, we compute the electric isovector
form factor. In this case, the regularization is necessary and
the second order terms in the FSD expansion have to be taken
into account.
 Contribution to the first order come from the terms of $W_2$
containing both $A_4$ and $\Omega$.
The second order includes terms from $W_1$ and $W_3$.

 In the first order FSD term we insert unity:
$\sum_m \psi^{\dagger}_m (x) \psi_m(x)$ and we get
identical overall isospin-space factor of the form:

$$ j^{em}_4 \sim \psi^{\dagger}_n \Omega \psi_m \
\psi^{\dagger}_m R \tau^3 R^{\dagger} \psi_n \ ,
\eqno(\EVCUR) $$

\noindent to which the Fierz identity (\APPAf) is applied.
Then, we express $\Omega$ in terms of the rotational
matrix $R$ using the identity:

$$ \Omega \equiv - i R {\dot R}^{\dagger} \equiv
-i {\dot R}^{\dagger} R
\eqno(\OMR) $$

\noindent and (\APPAg).

 Finally, from the quantization rule for the collective
coordinate $\Omega$ (\OMQ), where the moment of inertia $I$
has been computed numerically in [\GGGRW],
we arrive at the following expression for the
electric isovector form factor:

$$\openup3\jot
\eqalign{
G_E^{T=1} (q^2) = &\int d^3 x\ e^{-i \vec q \vec x}\
{-N_c\over 6I}
 \int d^3 y \ \Bigg\lbrace\ \sum_{n\in VAL\atop m\in ALL}
{\psi^{\dagger}_n(\vec x)\tau^A\psi_m(\vec x)\
\psi^{\dagger}_m(\vec y)\tau^A
\psi_n(\vec y) \over \epsilon_n - \epsilon_m}\ + \cr
&+ \ {1\over 4\sqrt\pi}
\sum_{m,n\in ALL}
\int_{1\over\Lambda^2}^{\infty} {d\tau\over\sqrt\tau} \,
\Bigg\lbrack {\epsilon_n e^{-\tau\epsilon_n^2} + \epsilon_m
e^{-\tau\epsilon_m^2}\over \epsilon_n + \epsilon_m } +
{1\over\tau} {e^{-\tau\epsilon_n^2} - e^{-\tau\epsilon_m^2}
\over \epsilon_n^2 - \epsilon_m^2}\ \Bigg\rbrack \ \times \cr
&\times \ \psi_n^{\dagger}(\vec x)\tau^A \psi_m(\vec x)\
\psi_m^{\dagger}(\vec y)\tau^A\psi_n(\vec y)\ \Bigg\rbrace \ , \cr
} \eqno(\ELVEC)  $$

 The formula (\ELVEC) for $q^2=0$ is equivalent to the quotient of the
moment of inertia by itself and:
$G^{T=1}_E(q^2=0) = 1$, as it should be.

 We proceed to compute the magnetic form factors. We start with
the isoscalar one.  In this case,
the contribution comes from the imaginary part of the action and
the regularization is not necessary. The quark current matrix $\hat Q$
contributes {\it via} the isoscalar term, $\sim {1\over 6}\,\,\hat 1$,
diagonal in the isospin space. Hence, the rotation matrices $R(t)$
cancel each other and, like for the electric scalar form factor,
there is no time ordering problem for the ${\cal D}R$ integration
in (\NEWGEN).
However, there is still a nonzero contribution from terms linear in
$\Omega$.
Starting from the reversed Hermicity of $A_{\mu}$
and the regularized (real)
action term (\WIcp) we find that contribution to the first order
FSD expansion comes from the anticommutator term:
$\gamma_4 \gamma_k A_k \, \{\Omega, R\hat Q R^{\dagger} \}
= {1\over 3} \gamma_4 \gamma_kA_k \Omega$.
It contributes to the electromagnetic current as follows:

$$ j_{em}^k(1) = {N_c \over 12 \sqrt{\pi} }
 \int_{1\over\Lambda^2}^{\infty} {d\tau\over\sqrt{\tau}} \sum_n
 {\rm e}^{-\tau\epsilon_n^2} \
\langle n \vert \gamma_4 \gamma_k \Omega
\vert n \rangle\ .
\eqno(\MSCURa) $$

 The second order FSD contribution comes from the terms
${1\over 3} A_k \gamma_4 \gamma_k \omega - {i\over 6}
[h, \gamma_4 \gamma_k A_k]$ (in $W_1$) and
$2\Omega\omega - i [h, \Omega]$  (in $W_3$).
Taking into account (\JFULL) and (\MSCURa)
together with (\RDEF) and (\OMQ) we obtain the space components
of the isoscalar electromagnetic current in Euclidean space-time in the
following form:

$$\openup3\jot
\eqalign{
j_k^{em} &= {- N_c\over 48\sqrt{\pi}\, I}
\int_{1\over\Lambda^2}^{\infty} {d\tau\over\sqrt{\tau}}
\sum_{m,n}
\ \Bigg\lbrack
{\epsilon_n {\rm e}^{-\tau\epsilon_n^2} +
\epsilon_m {\rm e}^{-\tau\epsilon_m^2} \over
\epsilon_n + \epsilon_m}
+ {1\over \tau}
{{\rm e}^{-\tau\epsilon_n^2} -
{\rm e}^{-\tau\epsilon_m^2} \over
\epsilon_n^2 - \epsilon_m^2}
\ \Bigg\rbrack \ \times \cr
&\times \ \int \, d^3y \ \langle N\vert \tau^A \vert N \rangle \
\psi_n^{\dagger}(\vec x) \gamma_4 \gamma_k \psi_m(\vec x)
\, \psi_m^{\dagger}(\vec y) \tau^A \psi_n(\vec y)
\  . \cr
} \eqno(\MSCUR)  $$

 Now, we have to perform transformation to the Minkowski space-time.
Also, one has once again to resolve (\FFDEFb), this time with respect
to the form factor $G_M^{T=0}$. To this end we apply (\APPAi) and
the result for the sea part of the magnetic isoscalar form factor
reads:

$$\openup3\jot
\eqalign{
G_M^{T=0} (q^2) \Bigg\vert_{sea} =
&\int d^3 x\ e^{-i \vec q \vec x}\
{-  N_c {\cal M}_N\over 24\sqrt{\pi} I}
 \int d^3 y \,\epsilon_{ijA} \  {q_j\over q^2} \ \times \cr
&\times \sum_{m,n\in ALL}
\int_{1\over\Lambda^2}^{\infty} {d\tau\over\sqrt\tau}
\ \Bigg\lbrack \, {\epsilon_n {\rm e}^{-\tau\epsilon_n^2} +
\epsilon_m
{\rm e}^{-\tau\epsilon_m^2}\over \epsilon_n + \epsilon_m } +
{1\over\tau} {{\rm e}^{-\tau\epsilon_n^2} -
{\rm e}^{-\tau\epsilon_m^2}
\over \epsilon_n^2 - \epsilon_m^2}\ \Bigg\rbrack \ \times \cr
&\times \ \psi_n^{\dagger}(\vec x)\gamma_0\gamma_i\psi_m(\vec x)\
\psi_m^{\dagger}(\vec y)\tau^A\psi_n(\vec y)\ . \cr
} \eqno(\MAGSCa)  $$

\noindent To obtain the full form factor we must add to (\MAGSCa)
the valence contribution and we must take the non-regularized limit:
$\Lambda^2\rightarrow\infty$. For the last step we use formulae
(\APPAa) to find that:

$$\openup3\jot
\eqalign{
&\lim_{\Lambda^2\to\infty} \int_{1/\Lambda^2}^{\infty}
{d\tau\over\sqrt{\tau}} \  \Bigg\lbrack
{\epsilon_n {\rm e}^{-\tau\epsilon_n^2} +
\epsilon_m
{\rm e}^{-\tau\epsilon_m^2}\over \epsilon_n + \epsilon_m } +
{1\over\tau} {{\rm e}^{-\tau\epsilon_n^2} -
{\rm e}^{-\tau\epsilon_m^2}
\over \epsilon_n^2 - \epsilon_m^2}\ \Bigg\rbrack = \cr
&= {-\sqrt{\pi}\over\epsilon_n - \epsilon_m}
\lbrack \ {\rm sign}(\epsilon_n) - {\rm sign}(\epsilon_m) \
\rbrack \ . \cr }
 \eqno(\MAGSCb) $$

 Finally, we get for the magnetic isoscalar form factor the following
expression:

$$\openup3\jot
\eqalign{
G_M^{T=0} (q^2) = &\int d^3 x\ e^{-i \vec q \vec x}\
{  N_c {\cal M}_N\over6I}
 \int d^3 y \,\epsilon_{ijA} \  {q_j\over q^2} \ \times \cr
&\times \Bigg\lbrace \sum_{n\in VAL\atop m\in ALL}
{\psi_n^{\dagger}(\vec x)\gamma_0\gamma_i\psi_m(\vec x)\
\psi_m^{\dagger}(\vec y)\tau^A\psi_n(\vec y) \over
\epsilon_n - \epsilon_m}  \cr
&+ \ {1\over 2} \ \sum_{n\in VAL\atop m\in ALL}
{\rm sign}(\epsilon_n) \
{\psi_n^{\dagger}(\vec x)\gamma_0\gamma_i\psi_m(\vec x)\
\psi_m^{\dagger}(\vec y)\tau^A\psi_n(\vec y) \over
\epsilon_n - \epsilon_m} \
\Bigg\rbrace \ , \cr
} \eqno(\MAGSC)  $$

\noindent where, as usually, summation over repeated indices $i, j, A$
is assumed. The first part gives the valence and the second part the sea
contribution to the form factor.

 For the magnetic isovector form factor the sea contribution
comes from the isotriplet part of the quark current matrix
$\hat Q$. Because of this we cannot get rid of the rotation
matrix $R(t)$ in the action.

 First, let us compute contribution from  the first order FSD expansion
(0--order in $\Omega$).
This term is the anticommutator $-{i\over 2} \{ h, \gamma_4
\gamma_k A_k R \tau^3 R^{\dagger} \}$  in (\WIb).
This implies for the electromagnetic current:

$$ j^{em}_k = {i N_c \over 4 \sqrt{\pi} } \sum_n
\int_{1\over\Lambda^2}^{\infty} {d\tau\over\sqrt{\tau}}
\epsilon_n \, {\rm e}^{-\tau\epsilon_n^2}
\langle n \vert \gamma_4 \gamma_k \tau^3
\vert n \rangle\ ,
\eqno(\MVCURp) $$

\noindent where we are still in the Euclidean space-time. Taking into
account (\APPAh) and $\gamma_0$ instead of $\gamma_4$ we get:

$$ j^{em}_k = {- N_c \over 8 \sqrt{\pi} } \
{\rm Tr} \, (R\tau^3 R^{\dagger} \tau^A \, )
\sum_n \int_{1\over\Lambda^2}^{\infty} {d\tau\over\sqrt{\tau}} \
\epsilon_n \, {\rm e}^{-\tau\epsilon_n^2} \
\langle n \vert \gamma_0 \gamma_k \tau^A
\vert n \rangle\ ,
\eqno(\MVCUR) $$

\noindent For the term (\MVCUR) regularization is necessary, as it
comes from the real part of the action and is infinite.

 For the linear order in $\Omega$ we have to take into account
non-commutativity of the isospin operator in
the collective variable (\OMQ) with the finite rotation
(Wigner) matrix.
This leads to the extra term, linear in the cranking parameter
The observation has been done in [\CORRa] in the context of
the axial coupling constant $g_A$, and the proper treatment
of such effect for the $g_A$ and the isovector magnetic
moment has been given in [\CORRb].

 Following the method of [\CORRb] we get the additional,
linear in $\Omega$ correction to the isovector electromagnetic
current in the form:

$$ j_{k}^{em\ A}(\Omega^1) =   N_c \
\lbrack \Omega_C ,\, {1\over 2} {\rm Tr} ( R^{\dagger}
\tau^A R \tau^B ) \rbrack
\sum_{m\in VAL\atop n\in ALL}
\int d^3y \
{\psi_n^{\dagger}({\vec x}) \gamma_0 \gamma_k \tau^B \  \psi_m({\vec x})
\psi_m^{\dagger}({\vec y}) \tau^C  \psi_n({\vec y})
\over\epsilon_n - \epsilon_m}  \ .
\eqno(\MVCURap) $$

\noindent Here, a remark concerning regularization of the
term (\MVCURap) is in order. This expression is numerically
finite and, as has been shown in [\CORRb], it is finite in the
gradient expansion. Hence, the regularization is not necessary.
Also, one should have in mind that the main contribution to the
isovector magnetic current comes from the valence part,
the sea contribution is relatively small and the formal regularization
of that contribution gives a small difference with
our result.

 To calculate the form factor we have to resolve the
equation (\FFDEFb) with respect to $G^{T=1}_M$ using
the spherical basis
in which the rotation matrices $R(t)$ are related to the Wigner
functions.
Our current (\MVCUR) has the general index structure of the form:
$j_k = X^A \, \epsilon_{kAi} \, q_i \, F(q)$.
Comparing this structure with eqs. (\FFDEFb,\FFSV), using (\APPAi)
we obtain the final expression for the magnetic isovector
form factor in the following form:

$$\openup3\jot
\eqalign{
G_M^{T=1} (q^2) = &\int d^3 x\ e^{-i \vec q \vec x}
\ {i N_c{\cal M}_N\over 3} \ \epsilon_{ijA} \ {q_j\over q^2}
\ \Bigg\lbrace \sum_{n\in VAL}
\psi_n^{\dagger}(\vec x)\gamma_0\gamma_i\tau^A\psi_n(\vec x)\ - \cr
&- {1\over 2\sqrt{\pi}} \sum_{n\in ALL}
\int_{1\over\Lambda^2}^{\infty} {d\tau\over\sqrt{\tau}}
\ \epsilon_n \
e^{-\tau\epsilon^2_n}\ \psi_n^{\dagger}(\vec x)\gamma_0\gamma_i\tau^A
\psi_n(\vec x)  \ +  \cr
&+ {1\over 6 I} \ \epsilon_{ABC}
\sum_{n>val\atop m\le val} \int d^3y \
{\psi_n^{\dagger}({\vec x}) \gamma_0 \gamma_i \, \tau^B
\  \psi_m({\vec x})
\psi_m^{\dagger}({\vec y}) \tau^C  \psi_n({\vec y})
\over\epsilon_n - \epsilon_m}
\Bigg\rbrace \ . \cr
} \eqno(\MAGVEC)  $$

 The proton and neutron form factors are defined, respectively, as sum
and difference of the isoscalar and isovector form factors:

$$\openup 3\jot
\eqalignno{
G^p_{E,M} &= {1\over 2}\ \lbrack \, G^{T=0}_{E,M} + G^{T=1}_{E,M}
\, \rbrack \ , &(\FFPNa) \cr
G^n_{E,M} &= {1\over 2}\ \lbrack \, G^{T=0}_{E,M} - G^{T=1}_{E,M}
\, \rbrack \ , &(\FFPNb) \cr
} $$

\noindent and the numerical results for them are presented in
Section 4.

\vskip\baselineskip
\vskip\baselineskip
\centerline{\bf 4. Numerical results}
\vskip\baselineskip

 To compute observables we use the finite quasi--discrete basis
and numerical method of Ripka and Kahana [\RIPKA] for solving
the eigenvalue problem. The Hamiltonian $h$ is taken in a spherical
box of a large radius $D$. The basis is made discrete by imposing
a boundary condition at $r=D$.
Also, it is made finite by restricting momenta of the basis states
to be smaller than the numerical cut-off $K_{max}$.
Both quantities have no physical meaning and the results
should not depend on them. The typical values used are
$D \sim 20/M$ and $K_{max} \sim 7 M$.

In addition, all checks concerning the numerical stability of the
solution with respect to varying box size and choice of the
numerical cut-off have been done and the actual calculation is
completely under control.

 The actual calculations are done by fixing the parameters in
meson sector in the well known way [\HEDGE] to have
$f_{\pi} = 93\ MeV$ and $m_{\pi} = 139.6\ MeV$.
This leaves the constituent quark mass $M$ as the only free
parameter.

 The proton and neutron electric and magnetic form factors
are displayed in figs.1--4. The theoretical curves resulting from the
model are given for the following four values the constituent quark mass:
$370, 400, 420$ and $450\ MeV$.
The magnetic form factors are normalized to the experimental
values of the corresponding magnetic moments at $q^2=0$.
With one exception of the neutron
electric form factor (fig.~2), all other form factors fit to the
experimental data quite well. The best fit is for the constituent quark mass
around $400$--$420\ MeV$.

As can be seen the only form factor which deviates from the experimental
data is the neutron electric form factor and this requires
some explanation. Obviously, this form factor is the most sensitive
for numerical errors.  According to the formula (\FFPNb)
the form factor has been calculated as a difference of the electric
isoscalar and electric isovector form factors that were the direct
output  of our code. Both form factors were of order $1.00$
and calculated by the code with rather high accuracy of about $1\%$.
However, the resulting neutron form factor has the proper (experimental)
value of order $0.04$, {\it i.e.} about $4\%$ of the value of its
components. This means that a small numerical error for one of the
components can be enhanced by a factor $50$.
Hence, small numerical errors together with the applied approximations
(like the $1/N_c$ approximation behind the model and
neglecting the boson--loop effects as well as higher order fermion
loops) are strongly magnified resulting in a considerable deviation from
the experimental data for momentum transfers above $100\ MeV/c$.

 As the next step, one can compute other electromagnetic observables:
the mean squared radii, the magnetic moments and the
nucleon--$\Delta$ splitting. The static nucleon properties, in particular
the charge radii and the magnetic moments can be obtained from the form
factors:

$$ \langle r^2 \rangle_{T=0,1} = - {dG_E^{T=0,1}\over
dq^2} \ {6\over G_E^{T=0,1}} \Bigg\vert_{q^2=0} \ ,
\eqno(\ELRAD) $$

$$ \mu^{T=0,1} = G_M^{T=0,1} (q^2) \Big\vert_{q^2=0} \ .
\eqno(\MAGMOM) $$

\vskip0.1\baselineskip

\noindent For the quark masses $370, 420$ and $450\
MeV$  and pion mass $m_{\pi}=140 MeV$ the calculated values are presented
in table 2. For comparison, in table 3 we give the theoretical values for
the same quantities but with the physical pion mass set to zero.

\vskip\baselineskip
\item{Table 2.} Nucleon observables computed with the physical
pion mass.
\vskip0.5\baselineskip

{\tenrm
\vbox{\offinterlineskip
\hrule height1pt
\halign{&\vrule width1pt#&
 \strut\quad\hfill#\hfill\quad
 &\vrule#&\quad\hfill#\hfil\quad
 &\vrule#&\quad\hfill#\hfil\quad
 &\vrule#&\quad\hfill#\hfil\quad
 &\vrule#&\quad\hfill#\hfil\quad
 &\vrule#&\quad\hfill#\hfil\quad
 &\vrule#&\quad\hfill#\hfil\quad
 &\vrule width1pt#&\quad\hfil#\hfil\quad\cr
height6pt&\omit&&\multispan{11}&&\omit&\cr
& &&\multispan{11} \hfill {\bf Constituent\ Quark\ Mass} \hfill &&
&\cr
height4pt&\omit&&\multispan{11}&&\omit&\cr
&\hfill {\bf Quantity}\hfill &&\multispan3 \hfill 370\ MeV\hfill
&&\multispan3
\hfill 420   MeV\hfill
&&\multispan3
\hfill 450  MeV\hfill
&&\hfill {\bf Exper.}\hfill &\cr
height3pt&\omit&&\multispan3 && \multispan3 &&
\multispan3 &&\omit&\cr
&\omit&&\multispan{11}\hrulefill&&\omit&\cr
height3pt&\omit&&\omit&&\omit&&\omit&&\omit&&\omit&&\omit&&\omit&\cr
&\hfill    &&\hfill total \hfill &&\hfill sea \hfill &&\hfill total
\hfill &&\hfill sea \hfill && \hfill total \hfill  &&\hfill sea \hfill  &&
&\cr %
height3pt&\omit&&\omit&&\omit&&\omit&&\omit&&\omit&&\omit&&\omit&\cr
\noalign{\hrule height1pt}
height3pt&\omit&&\omit&&\omit&&\omit&&\omit&&\omit&&\omit&&\omit&\cr

& \hfill $ <r^2>_{T=0} [fm^2] $ &&
0.63  &&
0.05  &&
0.52  &&
0.07  &&
0.48  &&
0.09  &&
0.62  &\cr
height3pt&\omit&&\omit&&\omit&&\omit&&\omit&&\omit&&\omit&&\omit&\cr
\noalign{\hrule}
height3pt&\omit&&\omit&&\omit&&\omit&&\omit&&\omit&&\omit&&\omit&\cr

& \hfill $ <r^2>_{T=1} [fm^2] $ &&
1.07  &&
0.33  &&
0.89  &&
0.41  &&
0.84  &&
0.45  &&
0.86 &\cr
height3pt&\omit&&\omit&&\omit&&\omit&&\omit&&\omit&&\omit&&\omit&\cr
\noalign{\hrule}
height3pt&\omit&&\omit&&\omit&&\omit&&\omit&&\omit&&\omit&&\omit&\cr

& \hfill $ <r^2>_p [fm^2] $ &&
 0.85  &&
 0.19  &&
 0.70  &&
 0.24  &&
 0.66  &&
 0.27  &&
 0.74  &\cr
height3pt&\omit&&\omit&&\omit&&\omit&&\omit&&\omit&&\omit&&\omit&\cr
\noalign{\hrule}
height3pt&\omit&&\omit&&\omit&&\omit&&\omit&&\omit&&\omit&&\omit&\cr

& \hfill $ <r^2>_n [fm^2] $ &&
--0.22   &&
--0.14   &&
--0.18   &&
--0.17   &&
--0.18   &&
--0.18   &&
--0.12   &\cr
height3pt&\omit&&\omit&&\omit&&\omit&&\omit&&\omit&&\omit&&\omit&\cr
\noalign{\hrule}
height3pt&\omit&&\omit&&\omit&&\omit&&\omit&&\omit&&\omit&&\omit&\cr

& \hfill $\mu_{T=0}\hfill [n.m.] $ &&
0.68  &&
0.09  &&
0.62  &&
0.03  &&
0.59  &&
0.05  &&
0.88  &\cr
height3pt&\omit&&\omit&&\omit&&\omit&&\omit&&\omit&&\omit&&\omit&\cr
\noalign{\hrule}
height3pt&\omit&&\omit&&\omit&&\omit&&\omit&&\omit&&\omit&&\omit&\cr

& \hfill $ \mu_{T=1}\hfill [n.m.] $ &&
 3.74 &&
 0.96 &&
 3.53 &&
 1.07 &&
 3.47 &&
 1.17 &&
 4.71        &\cr
height3pt&\omit&&\omit&&\omit&&\omit&&\omit&&\omit&&\omit&&\omit&\cr
\noalign{\hrule}
height3pt&\omit&&\omit&&\omit&&\omit&&\omit&&\omit&&\omit&&\omit&\cr

& \hfill $ \mu_p \hfill [n.m.] $ &&
  2.21 &&
  0.53 &&
  2.08 &&
  0.55 &&
  2.03 &&
  0.61 &&
  2.79        &\cr
height3pt&\omit&&\omit&&\omit&&\omit&&\omit&&\omit&&\omit&&\omit&\cr
\noalign{\hrule}
height3pt&\omit&&\omit&&\omit&&\omit&&\omit&&\omit&&\omit&&\omit&\cr
& \hfill $ \mu_n \hfill [n.m.] $ &&
 --1.53 &&
 --0.44 &&
 --1.46 &&
 --0.52 &&
 --1.44 &&
 --0.56 &&
 --1.91          &\cr
height3pt&\omit&&\omit&&\omit&&\omit&&\omit&&\omit&&\omit&&\omit&\cr
\noalign{\hrule}
height3pt&\omit&&\omit&&\omit&&\omit&&\omit&&\omit&&\omit&&\omit&\cr
& \hfill $ |\mu_p/\mu_n| \hfill $ &&
  1.44  &&
  ---   &&
  1.42  &&
  ---   &&
  1.41  &&
  ---   &&
  1.46           &\cr
height3pt&\omit&&\omit&&\omit&&\omit&&\omit&&\omit&&\omit&&\omit&\cr
\noalign{\hrule}
height3pt&\omit&&\omit&&\omit&&\omit&&\omit&&\omit&&\omit&&\omit&\cr
& \hfill $ M_{\Delta}-M_N [MeV] $ &&
 213  &&
 ---  &&
 280  &&
 ---  &&
 314  &&
 ---  &&
 294  &\cr
height3pt&\omit&&\omit&&\omit&&\omit&&\omit&&\omit&&\omit&&\omit&\cr
\noalign{\hrule}
height3pt&\omit&&\omit&&\omit&&\omit&&\omit&&\omit&&\omit&&\omit&\cr

& \hfill $ g_A \hfill $ &&
 1.32  &&
 0.15  &&
 1.29  &&
 0.20  &&
 1.29  &&
 0.23  &&
 1.26         &\cr
height3pt&\omit&&\omit&&\omit&&\omit&&\omit&&\omit&&\omit&&\omit&\cr
\noalign{\hrule height1pt} }}}
\vskip\baselineskip

The chiral limit $m_{\pi}\to 0$ mostly influences the isovector
charge radius. In fact, as illustrated in fig.~5, where the
isovector charge radius is plotted {\it vs.} the box size $D$,
it grows linearly with $D$ and diverges as $D\to\infty$.
Because of this that quantity (and the derivative quantities)
are not included in table 3.
The other quantity strongly influenced by the chiral limit
is the neutron electric form factor.
For the $m_{\pi}\to 0$ the discrepancy from the experiment is
larger by almost a factor two than in the case $m_{\pi}\neq 0$.
The other quantities differ in the chiral limit by about 30\%.
The comparison of values in the two tables indicate that
taking the physical pion mass gives us a best fit with
a much better agreement with the experimental data.
To be particular, in the
calculations with zero pion mass the electric observables suggest a
high value ($M\sim 450\ MeV$) while the magnetic ones indicate
$M\sim 370\ MeV$ which is not the case with physical pion mass. In addition,
we observe much larger contribution from the sea effects, about 50\% of the
total value.

\vskip\baselineskip
\item{Table 3.} Nucleon observables computed with the zero
pion mass.
\vskip0.5\baselineskip

{\twelvepoint

\vbox{\offinterlineskip
\hrule height1pt
\halign{&\vrule width1pt#&
 \strut\quad\hfil#\quad
 &\vrule#&\quad\hfil#\quad
 &\vrule#&\quad\hfil#\quad
 &\vrule#&\quad\hfil#\quad
 &\vrule#&\quad\hfil#\quad
 &\vrule#&\quad\hfil#\quad
 &\vrule#&\quad\hfil#\quad
 &\vrule#&\quad\hfil#\quad
 &\vrule width1pt#&\quad\hfil#\quad\cr
height6pt&\omit&&\multispan{11}&&\omit&\cr
& &&\multispan{11} \hfill {\bf Constituent\ Quark\ Mass} \hfill &&
&\cr
height4pt&\omit&&\multispan{11}&&\omit&\cr
&\hfill {\bf Quantity}\hfill &&\multispan3 \hfill 370\ MeV\hfill
&&\multispan3
\hfill 420   MeV\hfill
&&\multispan3
\hfill 450   MeV\hfill
&&\hfill {\bf Exper.} &\cr
height3pt&\omit&&\multispan3 && \multispan3 &&
\multispan3 &&\omit&\cr
&\omit&&\multispan{11}\hrulefill&&\omit&\cr
height3pt&\omit&&\omit&&\omit&&\omit&&\omit&&\omit&&\omit&&\omit&\cr
&\hfill    &&\hfill total &&\hfill sea &&\hfill total &&\hfill sea &&
\hfill total &&\hfill sea &&   &\cr
height3pt&\omit&&\omit&&\omit&&\omit&&\omit&&\omit&&\omit&&\omit&\cr
\noalign{\hrule height1pt}
height3pt&\omit&&\omit&&\omit&&\omit&&\omit&&\omit&&\omit&&\omit&\cr

& \hfill $ <r^2>_{T=0} \hfill [fm^2] $ &&
\hfill 0.88   &&
\hfill 0.20  &&
\hfill 0.66  &&
\hfill 0.26  &&
\hfill 0.61  &&
\hfill 0.23  &&
\hfill 0.62\hfill&\cr
height3pt&\omit&&\omit&&\omit&&\omit&&\omit&&\omit&&\omit&&\omit&\cr
\noalign{\hrule}
height3pt&\omit&&\omit&&\omit&&\omit&&\omit&&\omit&&\omit&&\omit&\cr

& \hfill $ \mu_{T=0} \hfill [n.m.] $ &&
\hfill 0.66  &&
\hfill 0.07  &&
\hfill 0.59  &&
\hfill 0.09  &&
\hfill 0.57  &&
\hfill 0.09  &&
\hfill 0.88\hfill
&\cr
height3pt&\omit&&\omit&&\omit&&\omit&&\omit&&\omit&&\omit&&\omit&\cr
\noalign{\hrule}
height3pt&\omit&&\omit&&\omit&&\omit&&\omit&&\omit&&\omit&&\omit&\cr

& \hfill $ \mu_{T=1} \hfill [n.m.] $ &&
\hfill 4.84   &&
\hfill 1.98  &&
\hfill 4.50  &&
\hfill 2.09  &&
\hfill 4.29  &&
\hfill 2.16  &&
\hfill 4.71 \hfill&\cr
height3pt&\omit&&\omit&&\omit&&\omit&&\omit&&\omit&&\omit&&\omit&\cr
\noalign{\hrule}
height3pt&\omit&&\omit&&\omit&&\omit&&\omit&&\omit&&\omit&&\omit&\cr

& \hfill $ \mu_p \hfill [n.m.] $ &&
\hfill   2.75  &&
\hfill   1.03  &&
\hfill   2.54  &&
\hfill   1.09  &&
\hfill   2.43  &&
\hfill   1.13  &&
\hfill   2.79 \hfill &\cr
height3pt&\omit&&\omit&&\omit&&\omit&&\omit&&\omit&&\omit&&\omit&\cr
\noalign{\hrule}
height3pt&\omit&&\omit&&\omit&&\omit&&\omit&&\omit&&\omit&&\omit&\cr

& \hfill $ \mu_n \hfill [n.m.] $ &&
\hfill --2.09  &&
\hfill --0.96  &&
\hfill --1.95  &&
\hfill --1.00  &&
\hfill --1.86  &&
\hfill --1.03  &&
\hfill --1.91 \hfill &\cr
height3pt&\omit&&\omit&&\omit&&\omit&&\omit&&\omit&&\omit&&\omit&\cr
\noalign{\hrule}
height3pt&\omit&&\omit&&\omit&&\omit&&\omit&&\omit&&\omit&&\omit&\cr

& \hfill $ |\mu_p/\mu_n| \hfill $ &&
\hfill 1.31 &&
\hfill ---  &&
\hfill 1.30  &&
\hfill ---  &&
\hfill 1.31  &&
\hfill ---  &&
\hfill 1.46 \hfill &\cr
height3pt&\omit&&\omit&&\omit&&\omit&&\omit&&\omit&&\omit&&\omit&\cr
\noalign{\hrule}
height3pt&\omit&&\omit&&\omit&&\omit&&\omit&&\omit&&\omit&&\omit&\cr

& \hfill $ M_{\Delta}-M_N \hfill [MeV] $ &&
\hfill 221\hfill &&
\hfill ---\hfill &&
\hfill 261\hfill &&
\hfill ---\hfill &&
\hfill 301\hfill &&
\hfill ---\hfill &&
\hfill 294\hfill &\cr
height3pt&\omit&&\omit&&\omit&&\omit&&\omit&&\omit&&\omit&&\omit&\cr
\noalign{\hrule}
height3pt&\omit&&\omit&&\omit&&\omit&&\omit&&\omit&&\omit&&\omit&\cr

& \hfill $ g_A \hfill [MeV] $ &&
\hfill 1.34 \hfill  &&
\hfill 0.24 \hfill &&
\hfill 1.29 \hfill &&
\hfill 0.15 \hfill &&
\hfill 1.25 \hfill &&
\hfill 0.00 \hfill &&
\hfill 1.26 \hfill &\cr
height3pt&\omit&&\omit&&\omit&&\omit&&\omit&&\omit&&\omit&&\omit&\cr
\noalign{\hrule}
\noalign{\hrule height1pt} }}}

\vskip\baselineskip

 The results of table 2  ($m_{\pi}=140$ MeV) again indicate the value $\sim
400$--$420$ MeV for the constituent quark mass, in
agreement with the conclusion drawn from the form factor curves.
The same has been suggested earlier [\GGGRW--\GGG], where
smaller number of observables has been taken into account.
With the exception of the neutron electric squared radii, to
which remarks similar to the case of the neutron electric form factor
are applicable, the contribution of the valence quarks is dominant. However,
the contribution of the Dirac sea is non-negligible, it lies within the
range 15 -- 40\%.

 One can notice, that the numerical results for
the nucleon--$\Delta$ mass splitting ($M_{\Delta}-M_N$),
the mean squared proton, isoscalar and isovector
electric radii and the axial coupling constant ($g_A$),
as well as the proton electric and magnetic and neutron
magnetic form factor, differ from the experimental data by no more then
about $\pm 5\%$. Finally, for the magnetic moments we have got results
20--25\% below their experimental values. Despite of this underestimation of
both magnetic moments we have found surprisingly good result for the
ratio $\mu_p/\mu_n$ which is far better than in other models.
The agreement with the experimental value is better than
3\% for the constituent quark mass 420 $MeV$.
So good result has been obtained neither in the Skyrme, nor
in the linear chiral model.

The isoscalar and isovector electric mean squared radii are
shown in figs.~6--7 as functions of the constituent quark
mass. The same plot but for the experimentally measured
quantities: the proton and neutron electric charge radii,
is given in fig.~8.
In these plots the valence and sea contributions are explicitly
given (dashed and dash-dotted lines, respectively). As could be expected,
for the isoscalar electric charge radius the valence part is dominant
(about 85\%), due to the fact that there is no contribution from
the terms linear in the collective coordinate $\Omega$.
This is not true for the isovector electric charge radius,
where the sea part contributes to about 45\% of the total value
(fig.~7). Also, this effect can be seen from the proton and neutron
charge radii (fig.~8) which are linear combinations of the
isoscalar and isovector ones.
For proton the sea contribution is about 30\%. However, for the
neutron charge radius the sea part is dominating and the valence
contribution is negligible.

 In addition, in fig.~9 we plot the magnetic moment density
distribution for proton and neutron for the constituent quark
mass $M = 420\ MeV$.
The sea contribution becomes non-negligible for distances
greater than $0.5\ fm$. Due to the relatively large tail contribution
to the magnetic moments the sea contribution to these quantities
is about 30\%.
If one is interested in the proton and neutron charge distribution
the reader is referred to ref. [\GGG].

Both the magnetic moments and the electric radii are rather sensitive to the
tail behaviour of the corresponding densities. In actual
calculations of these quantities the densities are integrated with extra
factors $r$ for the magnetic moments and $r^2$ for the radii that enhances
the role of the tail. This leads to the strong enhancement of numerical
errors and partially it explains the deviations.

\vskip\baselineskip
\vskip\baselineskip
\centerline{\bf 5. Summary and conclusions}
\vskip\baselineskip

Our numerical results support the view that chiral quark soliton model
of the Nambu--Jona-Lasinio type offers relatively simple but
quite successful description of some low-energy QCD phenomena and,
in particular, of the electromagnetic properties of nucleons.
For the first time the magnetic form factors are calculated.
We have obtained quite
good results for the electromagnetic form factors, the mean squared radii,
the magnetic moments and the nucleon--$\Delta$ splitting
reported in figs.~1--4 and tables 2, 3.

 Using $f_{\pi} = 93\ MeV$ and $m_{\pi} = 139.6\ MeV$ and a
constituent quark mass of $M = 420\ MeV$ the isoscalar and
isovector electric radii are reproduced within 15\%. The magnetic moments
are underestimated by about 25\%, however, their ratio $\mu_p/\mu_n$ is
almost perfectly reproduced. The $q$--dependence
of $G^p_E(q^2)$, $G^p_M(q^2)$ and $G^n_M(q^2)$ is very well reproduced.
The neutron form factor $G^n_E(q^2)$ is by a factor of two
too large for $q^2 > 100\ MeV^2$. One should note, however,
that $G^n_E(q^2)$ is more than an order of magnitude smaller
than $G^p_E(q^2)$  and as such it is extremely sensitive
to both the model approximations and the numerical techniques used.
 Altogether it is fair to say that the NJL in $SU(2)$
reproduces reasonably well the electromagnetic properties of
the nucleon.
Here, it turns out that the agreement
is noticeably worse if the chiral limit, $m_{\pi}\to 0$, is
used. This can easily be understood since the pion mass
determines the asymptotic behaviour of the pion field.
Altogether we conclude that the bosonized NJL model
chiral quark soliton model based on a bosonized NJL type lagrangean is quite
appropriate for the evaluation of nucleonic electromagnetic properties.

\vskip0.5\baselineskip

\noindent{\bf Acknowledgement:}

{\it We would like to thank Victor Petrov
and Pavel Pobylitsa for numerous helpful discussions.
The project has been partially supported by the BMFT,
DFG and COSY (J\"ulich).
Also, we are greatly indebted for financial support to
the Bulgarian Science Foundation $\Phi$--32
(CVC) and to the Polish Committee for
Scientific Research,  Projects KBN nr 2 P302 157 04
and 2 0091 91 01 (AZG).}

\vskip\baselineskip

{\bf References}
\vskip\baselineskip

\def\SKYRref{\item{[\SKYR]} T.H.Skyrme
{\it Nucl.Phys.} {\bf 31} (1962) 556.}

\def\NJLref{\item{[\NJL]} J.Nambu, G.Jona-Lasinio, {\it Phys.Rev.}
{\bf 122} (1961) 345.}

\def\HEDGEref{\item{[\HEDGE]} Th.Meissner, K.Goeke, F.Gr\"ummer, {\it
Phys.Lett.} {\bf 227B} (1989) 296; {\it Ann.Phys.} {\bf 202} (1990) 297.}

\def\MESSECaref{\item{[\MESSECa]} D.Ebert, M.K.Volkov,
{\it Z.Phys.} {\bf C16} (1983) 205.        }

\def\MESSECbref{\item{[\MESSECb]} V.Bernard {\it et al}
{\it Nucl.Phys.} {\bf A412} (1984) 349.   }

\def\BARSECaref{\item{[\BARSECa]} Th.Meissner,  K.Goeke, {\it
Nucl.Phys.} {\bf A524} (1991) 719. }

\def\RWref{\item{[\RW]}H.Reinhardt, R.W\"unsch, {\it Phys.Lett.}  {\bf 215B}
(1989) 825.}

\def\DPref{\item{[\DP]} D.Dyakonov, V.Petrov,
{\it Phys.Lett.} {\bf 147B} (1984) 351; {\it Nucl.Phys.} {\bf B272} (1986)
457.}

\def\DPPref{\item{[\DPP]} D.Dyakonov, V.Petrov, P.Pobylitsa,
{\it Nucl.Phys.} {\bf B306} (1988) 809. }

\def\BALLref{\item{[\BALL]} R.Ball, {\it Phys.Rep.}
{\bf 182} (1989) 1.}

\def\SRALref{\item{[\SRAL]} M.Schaden, H.Reinhardt, P.Amundsen,
M.Lavelle, {\it Nucl.Phys.} {\bf B339} (1990) 595. }

\def\EGUCHIref{\item{[\EGUCHI]} T.Eguchi, H.Sugawara,
{\it Phys.Rev.} {\bf D10} (1974) 4257;
T.Eguchi, {\it Phys.Rev.} {\bf D14} (1976) 2755.  }

\def\GGGRWref{\item{[\GGGRW]} K.Goeke, A.Z.G\'orski, F.Gr\"ummer,
Th.Meissner, H.Reinhardt, R.W\"unsch,
{\it Phys.Lett.} {\bf B256} (1991) 321. }

\def\GGGref{\item{[\GGG]} A.Z.G\'orski, K.Goeke, F.Gr\"ummer,
{\it Phys.Lett.} {\bf B278} (1992) 24. }

\def\SCHWref{\item{[\SCHW]} J.Schwinger, {\it Phys.Rev.}
{\bf 82} (1951) 664.  }

\def\RIPKAref{\item{[\RIPKA]} S.Kahana, G.Ripka, {\it Nucl.Phys.}
{\bf A419} (1984) 462. }

\def\WAKAMref{\item{[\WAKAM]} M.Wakamatsu, H.Yoshiki,
{\it Nucl.Phys.} {\bf A524} (1991) 561. }

\def\CORRaref{\item{[\CORRa]} M.Wakamatsu, T.Watabe,
{\it Phys.Lett.} {\bf B312} (1993) 184. }

\def\CORRbref{\item{[\CORRb]} Ch.Christov, K.Goeke,
V.Yu.Petrov, P.V.Pobylitsa, M.Wakamatsu, T.Watabe,
preprint RUB-TPII-53/93, Nov 1993
(submitted to: {\it Phys.Lett.} {\bf B}).  }

\SKYRref
\NJLref
\EGUCHIref
\DPref
\DPPref
\RWref
\HEDGEref
\MESSECaref
\MESSECbref
\BARSECaref
\BALLref
\SRALref
\GGGRWref
\WAKAMref
\GGGref
\RIPKAref
\SCHWref
\CORRaref
\CORRbref

\newpage

\centerline{\bf Appendix}
\vskip\baselineskip

In Euclidean space we use the metric tensor's signature transformed from $(+
- - -)$ to $(- - - -)$ ({\it i.e.} the Euclidean metric tensor
$g_{\mu\nu}^{E}\equiv -\delta_{\mu\nu}$)
and the general formulae to perform the transformation from Minkowski to
Euclidean space are:

$$ \openup 1\jot \eqalign{
&a_4^E = +i a_0^M, \quad\quad a^4_E = - i a^0_M , \quad\quad\
 a_k^E = +a_k^M ,  \cr
&\gamma^E_4 = + i \gamma_0^M ,  \quad\quad\
 \gamma_k^E = +\gamma_k^M , \cr
&\{ \gamma^E_{\mu}, \gamma^E_{\nu} \} = -2 \delta_{\mu\nu} , \quad\quad
 \{ \gamma_{\mu}^E, \gamma_5^E \} = 0 \ , \cr
&( \gamma_{\mu}^E )^{\dagger} = - \gamma_{\mu}^E , \quad\quad
 (\gamma_5^E)^{\dagger} = + \gamma^E_5 = \gamma_5^M \ . \cr
}
\eqno(\EUCLID)  $$
\vskip0.5\baselineskip

Here, the simple substitution rule for the
limits of basic $\tau$-integrals are
given to calculate the non-regularized formulae for observables:

$$ \openup2\jot \eqalignno{
\lim_{\Lambda^2\to\infty} \int_{1/\Lambda^2}^{\infty}
{d\tau\over\tau^{1/2}} \
\alpha \ e^{-\tau\alpha^2} &\to \sqrt{\pi} \ \sign(\alpha) \ ,
&(\APPAaa) \cr
\lim_{\Lambda^2\to\infty} \int_{1/\Lambda^2}^{\infty}
{d\tau\over\tau^{3/2}} \
e^{-\tau\alpha^2} &\to -2 \pi \, \vert\alpha\vert \ ,
&(\APPAab) \cr
} $$

\noindent where $\alpha$ is a real number. In addition,
we can express the
$\sign(x)$ function in terms of the step function as:

$$ \sign(x) \equiv \Theta(x) - \Theta(-x) \equiv 1 - 2 \Theta(-x)
\ . \eqno(\APPAb) $$

 To obtain the formulae (\WI) we have applied simple identity for
the time derivative of the operator $R\hat Q R^{\dagger}$,
implied by (\RDEF):

$$ \partial_4 (R \hat Q R^{\dagger}) = i \, [ R\hat Q R^{\dagger},
\Omega ]
\ . \eqno(\APPAc) $$

\noindent Also, one should have in mind that our potentials
$A_{\mu}(x)$ are time-independent and we can use identity:

$$ \partial_4 A_4 \equiv \partial_4 A_k \equiv 0
\ . \eqno(\APPAd) $$

 For the second order FSD expansion it is useful to apply the
Fierz identity. If the matrices $\hat A, \hat B$:
$\hat A \equiv {1\over 2} A^A \tau^A, \
\hat B \equiv {1\over 2} B^A \tau^A$, then we get:

$$ {\rm Tr}\, ( \hat A \tau^A )\ {\rm Tr}\, ( \hat B \tau^A ) =
2\, {\rm Tr}\, (\hat A \hat B)
\ . \eqno(\APPAe) $$

\noindent This can be applied to our overall isospin factor in
the electromagnetic current as follows:

$$\openup 2\jot
\eqalign{
&\sum_{m,n}\int d\Omega_{\vec x} d\Omega_{\vec y} \psi^{\dagger}_n(\vec y)
\Omega \psi_m(\vec
y) \psi^{\dagger}_m(\vec x) R^{\dagger} \tau^3 R \psi_n(\vec x) = \cr
&\sum_{m,n} {1\over 4} {\rm Tr} ( \Omega \tau^A ) \ {\rm Tr}
( R^{\dagger} \tau^3 R \tau^B ) \
\int d\Omega_{\vec x} d\Omega_{\vec y}\psi^{\dagger}_n(\vec y) \tau^A
\psi_m(\vec y) \psi^{\dagger}_m(\vec x) \tau^B \psi_n(\vec x) = \cr
&\sum_{m,n} {1\over 12} {\rm Tr} ( \Omega \tau^A ) \ {\rm Tr}
( R^{\dagger} \tau^3 R \tau^A ) \
\int d\Omega_{\vec x} d\Omega_{\vec y}\psi^{\dagger}_n(\vec y) \tau^B
\psi_m(\vec y) \psi^{\dagger}_m(\vec x) \tau^B \psi_n(\vec x) = \cr
&\sum_{m,n} {1\over 6} {\rm Tr} ( \Omega R^{\dagger} \tau^3 R ) \
\int d\Omega_{\vec x} d\Omega_{\vec y}\psi^{\dagger}_n(\vec y) \tau^A
\psi_m(\vec y) \psi^{\dagger}_m(\vec x) \tau^A \psi_n(\vec x) \ . \cr
}  \eqno(\APPAf) $$

 Further simplification is done using the identity:

$$ {\rm Tr} \, (\Omega R \tau^3 R^{\dagger} ) =
-i \ {\rm Tr} \, (R {\dot R}^{\dagger} R \tau^3 R^{\dagger} )
= -i \ {\rm Tr} \, ({\dot R}^{\dagger} R \tau^3 ) = \Omega^3 \ ,
\eqno(\APPAg) $$

\noindent where we have used (\OMR) and the obvious identity:

$$ \hat A \equiv {1\over 2}\ {\rm Tr}\, (\hat A \tau^A) \ \tau^A \ .
\eqno(\APPAh) $$

 To resolve the magnetic form factor from the right hand side of eq.
(\FFDEFb) we use the identity:

$$ 1 \equiv {1\over 2}\ \epsilon_{mBl}\ \epsilon_{mBj}\
{q_l \, q_j\over q^2} \ .
\eqno(\APPAi) $$

\noindent This can be applied to our expression for the current
(\MVCUR).


\newpage

\centerline{\bf Figure captions}

\vskip\baselineskip

\item{Fig. 1.} The proton electric form factor for the
momentum transfers below $1\ GeV$.
\vskip0.3\baselineskip

\item{Fig. 2.} The neutron electric form factor for the
momentum transfers below $1\ GeV$.
\vskip0.3\baselineskip

\item{Fig. 3.} The proton magnetic form factor normalized to one
at $q^2=0$ for the momentum transfers below $1\ GeV$.
The normalization factor can be extracted from table 2.
\vskip0.3\baselineskip

\item{Fig. 4.} The neutron magnetic form factor normalized to one
at $q^2=0$ for the momentum transfers below $1\ GeV$.
The normalization factor can be extracted from table 2.
\vskip0.3\baselineskip

\item{Fig. 5.} The isovector charge radius as a function of
the box size $D$ for $m_{\pi}=0$ and $m_{\pi}=139.6\ MeV$.
\vskip0.3\baselineskip

\item{Fig. 6.} The isoscalar electric charge radius as a function of
the constituent quark mass $M$. The valence and sea parts are marked
by the dashed and dashed-dotted lines.
\vskip0.3\baselineskip

\item{Fig. 7.} The isovector electric charge radius as a function of
the constituent quark mass $M$. The valence and sea parts are marked
by the dashed and dashed-dotted lines.
\vskip0.3\baselineskip

\item{Fig. 8.} The electric charge radii of proton and neutron
as functions of the constituent quark mass $M$. The valence and sea
parts are marked by the dashed lines.
\vskip0.3\baselineskip

\item{Fig. 9.} The magnetic moment
density of proton and neuron
for the constituent quark mass $M =$  $420\ MeV$.


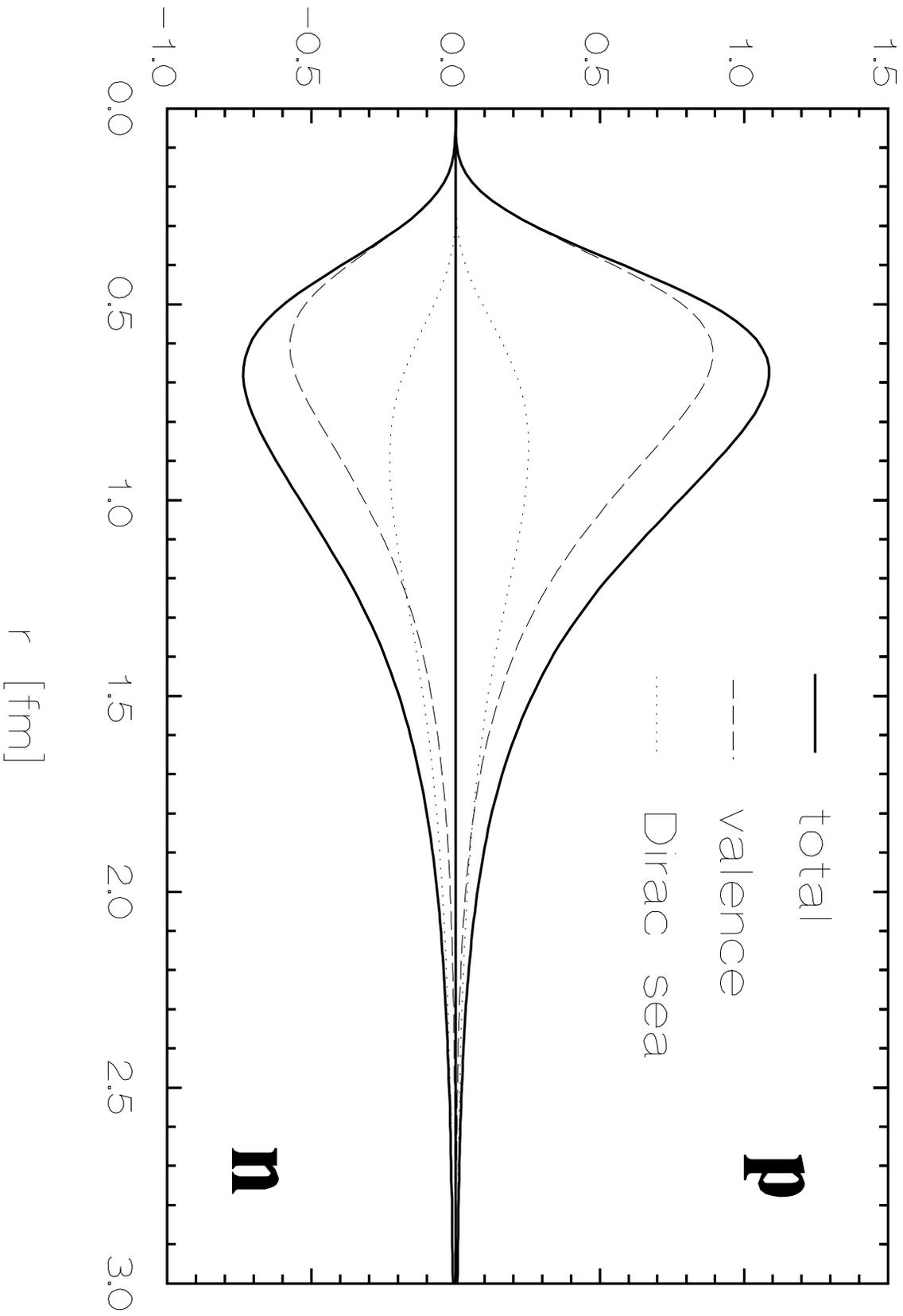
\bye